\documentclass[prd,preprint,showpacs,groupedaddress]{revtex4-1}
\usepackage{amsmath}
\usepackage{amsfonts}
\usepackage{amssymb}
\usepackage{geometry}
\usepackage{graphicx}
\usepackage{natbib}
\usepackage[english]{babel}
\usepackage{graphicx}
\usepackage{subfigure}
\usepackage{epstopdf}
\usepackage{caption}
\usepackage{multirow}
\usepackage{indentfirst}
\usepackage{siunitx}
\usepackage{diagbox}
\usepackage{mathrsfs}
\usepackage{booktabs}
\usepackage[colorlinks,linkcolor=blue,anchorcolor=,citecolor=blue]{hyperref}
\usepackage{amsthm}
\usepackage{float}
\numberwithin{equation}{section}
\numberwithin{figure}{section}
\numberwithin{table}{section}

\begin{document}
	\title{Decoding quantum gravity information with black hole accretion disk}
	\author{Lei You}
	\affiliation{Lanzhou Center for Theoretical Physics, Key Laboratory of Theoretical Physics of Gansu Province, Lanzhou University, Lanzhou, Gansu 730000, China}
	
	\author{Yu-Hang Feng}
	\affiliation{Lanzhou Center for Theoretical Physics, Key Laboratory of Theoretical Physics of Gansu Province, Lanzhou University, Lanzhou, Gansu 730000, China}
	
	\author{Rui-Bo Wang}
	\affiliation{Lanzhou Center for Theoretical Physics, Key Laboratory of Theoretical Physics of Gansu Province, Lanzhou University, Lanzhou, Gansu 730000, China}
	
	\author{Jian-Bo Deng}
	\affiliation{Lanzhou Center for Theoretical Physics, Key Laboratory of Theoretical Physics of Gansu Province, Lanzhou University, Lanzhou, Gansu 730000, China}
	
	\author{Xian-Ru Hu}
	\email[Email: ]{huxianru@lzu.edu.cn}
	\affiliation{Lanzhou Center for Theoretical Physics, Key Laboratory of Theoretical Physics of Gansu Province, Lanzhou University, Lanzhou, Gansu 730000, China}
	
	\date{\today}
	\begin{abstract}
		The combination of Loop Quantum Gravity theory with the classical gravitational collapse model has effectively addressed the singularity problem of black holes and predicted the emergence of white holes in the late stages of collapse. The quantum extension of Kruskal spacetime suggests that the appearance of white holes may carry information from companion black holes in the universe earlier than ours. Photons emitted from the accretion disk of companion black holes will enter the companion black hole, traverse through quantum regions from the white hole to our universe, and produce imaging of accretion disk carrying quantum gravity information. In our work, we have obtained the accretion disk images of black hole from a universe earlier than ours, transported by a white hole within our universe, along with the positions and widths of these images exactly. Remarkably, behaviours of white hole and black hole imaging are similar in photon sphere and contrary to some cases of outside. This will provide valuable references for astronomical observations to validate quantum gravity theory.
	\end{abstract}
	
	\maketitle
	\section{Introduction}
	    General relativity(GR) has provided an excellent description of gravity, yet the problem of singularities \cite{QQD1,QQD2,QQD3} remains an unresolved challenge. Theoretical physicists have devoted considerable effort to this issue, with the quantization of gravity emerging as the mainstream approach. Consequently, numerous quantum gravity(QG) theories have been proposed \cite{QD1,QD2,QD3,QD4,QD5,QD6}. However, experimental validation is essential for these theories. In 2019, the Event Horizon Telescope(EHT) captured the first image of black hole(BH) \cite{M87}, thereby confirming the prediction made by traditional gravitational theory. For QG, researchers also aspire to validate their theories through observations of BH images.
	    
	    Loop quantum gravity(LQG) is a QG theory aimed at reconciling GR with quantum mechanics. Unlike string theory, in LQG, spacetime is quantized at the smallest scale known as the Planck scale. This quantization of geometric shapes leads to a discrete structure of spacetime, where space is no longer continuous but composed of discrete units or loops \cite{LQG1,LQG2,LQG3,LQG4,LQG5,LQG6,LQG7,LQG8}. This crucial feature avoids the singularity problem inherent in GR. When collapsing matter reaches Planck-scale density near a BH singularity, quantum effects become dominant. LQG predicts a quantum bounce instead of singularities with infinite density and curvature, where collapsing matter rebounds due to quantum repulsion. At the Planck scale, the discrete nature of space prohibits the existence of singularity. This structure of spacetime in LQG acts as a barrier, preventing collapse and triggering a bounce, resulting in the formation of a quantum bridge or new regions of spacetime~\cite{C1,C2,C3,C4,C5,C6}. Thus, LQG demonstrates a significant strength in addressing the problem of BH singularities.
	    
\begin{figure}[htbp]
	    	\centering
	    	\includegraphics[width=0.7\textwidth]{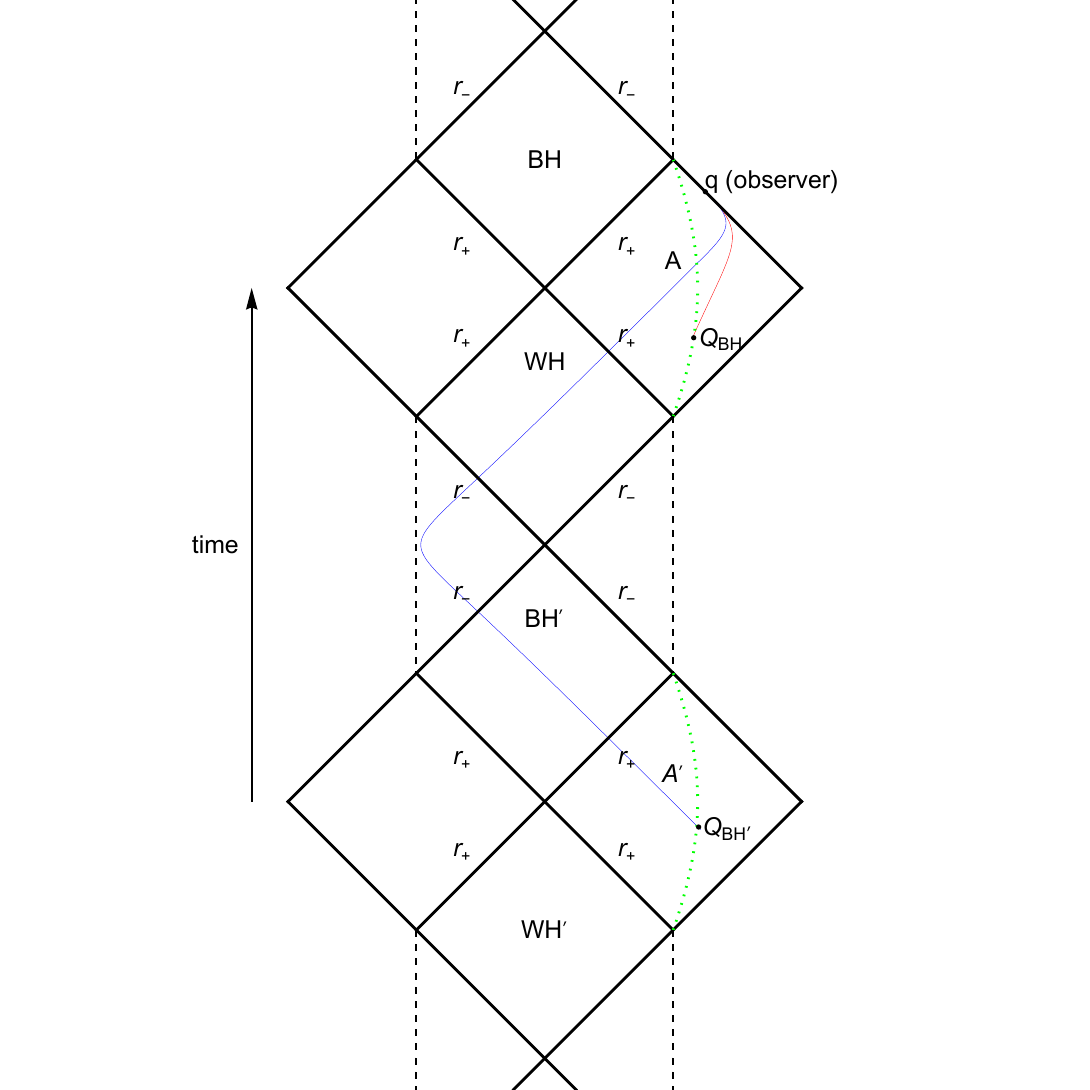}
	    	\caption{The Penrose diagram of the quantum extended BH spacetime. The two green dotted lines represent the orbits of the accretion disks around the BH within our universe (A) and the BH$^\prime$ in a universe earlier than ours (A$^\prime$), respectively. The red line and the blue line respectively represent the trajectories of photons emitted from point $Q_{\rm{BH}}$ in location A and point $Q_{\rm{BH}^{\prime}}$ in location A$^\prime$, reaching an observer at infinity denoted as point $q$.}\label{prs}
	    \end{figure}

	    In recent years, there has been a significant progress in combining classical gravitational collapse models (OS model) \cite{OS} with the LQG theory to study the gravitational collapse of spherically symmetric dust matter, considering quantum effects~ \cite{OS1,OS2,OS3,OS4,OS5,OS6,OS7,OS8,OS9}. Recently, Lewandowski, Ma, and Yang have obtained a new LQG BH solution by modifying the Schwarzschild black hole(SBH) \cite{BHWH}. In their model, during the initial stages of collapse, the external spacetime of the dust ball is described by the quantum-corrected SBH. In the late stages, the dust ball transitions from collapse to expansion, leading to a transition from a BH to a white hole(WH). Concurrently, considering the ideas proposed by Ashtekar et al. regarding companion BHs \cite{BX1,BX2,BX3}, photons may have the opportunity to enter from the companion BH, traverse through highly quantum region, and emerge from the WH in our universe. 
	    This presents a promising opportunity for observing quantum gravity effects.
	    As shown in Fig.~\ref{prs}, point $Q_{\rm{BH}^{\prime}}$ represents a particle in the accretion disk orbiting around a BH$^\prime$ located in a universe earlier than ours. Photons emitted from point $Q_{\rm{BH}^{\prime}}$ enter the BH$^\prime$ and exit from a WH within our universe, eventually reaching an observer at infinity, thus forming an additional accretion disk image point. Point $Q_{\rm{BH}}$ represents a particle in the accretion disk orbiting around a BH within our universe. Photons emitted from this point reach an observer at infinity, forming a typical accretion disk image point.
	    In~\cite{YJS,BX}, Zhang and Yang investigate this particular observational effect. Building upon their work, we further study the imaging of accretion disk caused by WH. This research aims to expand our understanding of the fundamental properties of BHs and the QG effects controlling their behavior, providing a new theoretical foundation for experimental observations of QG.
	    
	    Our exposition is as follows: In Sec.~\ref{sec2}, we investigate how photons propagate in quantum modified spacetime. In Sec.~\ref{sec3}, we establish an observer coordinate system and depict the imaging of accretion disks caused by BH and WH respectively. Furthermore, employing the Novikov-Thorne model, we generate intensity images of accretion disk emissions. Finally, in Sec.~\ref{sec4}, we summarize our findings.
	    \newpage
	
	\section{NULL GEODESICS IN THE QUANTUM MODIFIED SPACETIME}\label{sec2}
		For a static spherically symmetric quantum black hole (QBH) solution, the line element is~\cite{BHWH}
	\begin{equation}
		ds^2=-f(r)dt^2+f(r)^{-1}dr^2+r^2(d\theta^2+\sin^2\theta d\phi^2),
	\end{equation}
	    with
	\begin{equation}
	    f(r)=1-\frac{2M}{r}+\frac{\beta M^{2}}{r^{4}},\quad\beta=16\sqrt{3}\pi\gamma^{3}\ell_{p}^{2},
	\end{equation}
	    where $\ell_{p}=\sqrt{\frac{\hbar G}{c^3}}$ is Planck length and $\gamma$ is the Barbero-Immirzi parameter~\cite{GM1,GM2}. We adopt natural units ($c=G=\hbar=1$), resulting in $\beta=16\sqrt{3}\pi\gamma^3\approx1.16633$. To ensure that $f(r)=0$ possesses real roots, $M$ has a minimum value of $M_{\mathrm{min}}=\frac{16\gamma\sqrt{\pi\gamma}}{3\sqrt{3}}\approx0.8314$.  The horizon radius $r_{\pm}$ of this BH can be expressed as	    
	    \begin{equation}
	    \begin{split}
	    	r_{\pm}&=\frac{1}{6}(3M+\sqrt{3}\sqrt{\left( 6\sigma \right) ^{\small{\frac{1}{3}}}+3M^2+\frac{2\times 6^{\small{\frac{2}{3}}}M^2\beta}{\left( \sigma \right) ^{\small{\frac{1}{3}}}}})\\
	    	&\pm\frac{1}{6}\sqrt{3}\sqrt{6M^2-\frac{2\times 6^{\small{\frac{2}{3}}}M^2\beta}{\left( \sigma \right) ^{\small{\frac{1}{3}}}}-\left( 6\sigma \right) ^{\small{\frac{1}{3}}}+\frac{6\sqrt{3}M^3}{\sqrt{\left( 6\sigma \right) ^{\small{\frac{1}{3}}}+3M^2+\frac{2\times 6^{\small{\frac{2}{3}}}M^2\beta}{\left( \sigma \right) ^{\small{\frac{1}{3}}}}}}},
	    \end{split}
	    \end{equation}
	    where $\xi =\sqrt{3}\sqrt{M^6\left( 27M^2-16\beta \right) \beta ^2}$ and $\sigma =9M^4\beta +\xi$.

	    Now, let us derive the equations of motion for the particle. A particle's Lagrangian is
	    \begin{equation}\label{eq2_4}
	    	\mathcal{L}=\frac{1}{2}g_{\mu\nu}\dot{x}^{\mu}\dot{x}^{\nu}=\frac{1}{2}\left(-f\left(r\right)\dot{t}^2+\frac{1}{f\left(r\right)}\dot{r}^2+r^2\dot{\theta}^2+r^{2}\sin^{2}{\theta}\dot{\phi}^{2}\right),
	    \end{equation}
	    where $\dot{x}^{\mu}=\frac{dx^{\mu}}{d\lambda}$. For photon, $\lambda$ is affine parameter and for time-like particle, $\lambda$ is proper time $\tau$. There are two Killing vector fields in static spherically symmetric spacetime $\frac{\partial}{\partial t}$ and $\frac{\partial}{\partial \phi}$, so particle has its two conserved quantities
	    \begin{equation}\label{eq2_5}
	    	E=-\frac{\partial \mathcal{L}}{\partial \dot{t}}=f\left(r\right)\dot{t},
	    \end{equation}
	    \begin{equation}\label{eq2_6}
	    	L=\frac{\partial \mathcal{L}}{\partial \dot{\phi}}=r^{2}\dot{\phi},
	    \end{equation}
	    where $E$ is particle's energy and $L$ is angular momentum. Please notice that we have chosen $\theta=\frac{\pi}{2}$, which means that particle always moves on equatorial plane. Impact parameter $b$ is defined as $b:=\frac{|L|}{E}$.
	    
	    In consideration of the Lagrangian of photons being zero, by simultaneously solving these three equations, we obtain the equation of motion for photons on the equatorial plane as
	\begin{equation}\label{eq2_16}
		G\left(u\right):=\left(\frac{du}{d\phi}\right)^{2}=\frac{1}{b^{2}}-u^{2}f\left(\frac{1}{u}\right).
	\end{equation}
		By solving the system of equations $G\left( u \right)=G^{\prime}\left( u \right)=0$, we obtain the photon sphere radius $r_{ph}$ and its corresponding impact parameter $b_c$ as \cite{GZQ}
		\begin{equation}
			r_{ph}\simeq3M-\frac{\beta}{9M},
		\end{equation}
		\begin{equation}
			b_{c}=3\sqrt{3}M-\frac{\beta}{6\sqrt{3}M}+{\mathcal O}(\beta^{2}).
			\end{equation}
			
\begin{figure}[htbp]
				\centering
				\includegraphics[width=1\textwidth]{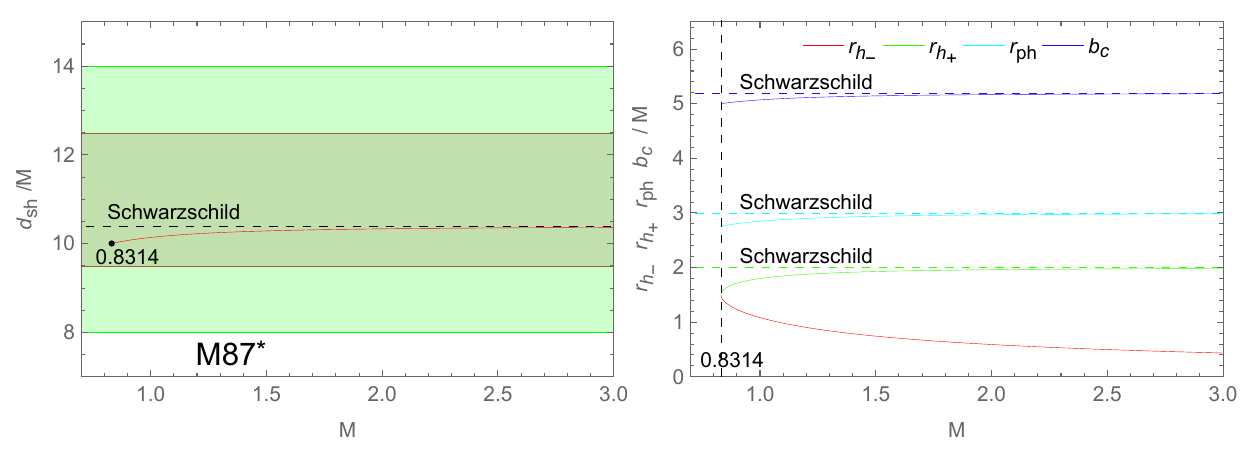}
				\caption{The left side is the variation of the QBH's shadow diameter with respect to $M$. The brown and
					orange shaded regions represent the regions of $1\sigma$ and $2\sigma$ confidence intervals, respectively, with
					respect to the M87* observations. On the right, the variations of $r_{h_{-}}$, $r_{h_{+}}$, $r_{ph}$, and $b_c$ with respect to $M$ are depicted.}
					\label{M87}
			\end{figure}

			To integrate with observational reality, we employ observational data from the M87* BH to examine the range of possible values for the mass $M$ of the QBH. As widely acknowledged, the angular size of the M87* shadow is denoted by $\delta~=~(42\pm3)~\mu\text{as}$, its distance by $D=16.8_{-0.7}^{+0.8}$, and its mass by $M=(6.5\pm0.9)\times10^9M_\odot$. By utilizing this data, the shadow diameter $d_{M87^{*}}=\frac{D\delta}{M}\simeq11.0\pm1.5$ of the BH can be computed~\cite{DM871,DM872}. As shown in Fig.~\ref{M87}, the $M$ can be constrained as $M\geqslant 0.8314$. After constraining the range of $M$, we plot the variations of $r_{h_{-}}$, $r_{h_{+}}$, $r_{ph}$, and $b_c$ with respect to $M$. It can be observed that as $M$ increases, $r_{h_{+}}$ and $r_{h_{-}}$ gradually separate, and $r_{h_{+}}$, $r_{ph}$, and $b_c$ approach the characteristics of a SBH. In subsequent calculations, we set $M=1.0859$.
	\begin{figure}[htbp]
		\centering
		\includegraphics[width=1\textwidth]{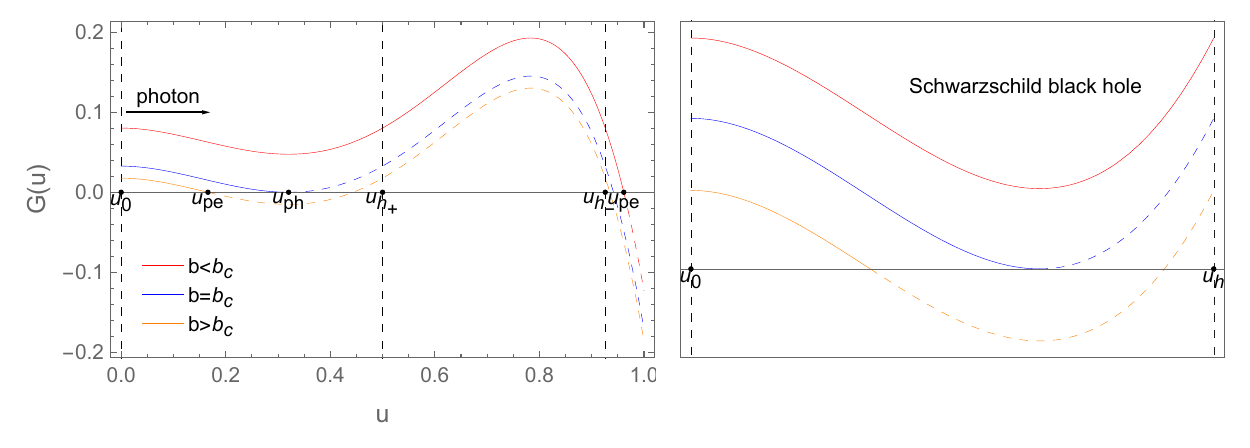}
		\caption{Functions $G\left(u\right)$ of $u$. On the left and right are respectively the QBH and SBH spacetimes. $u_{0}=\frac{1}{r_{\infty}}$, $u_{ph}=\frac{1}{r_{ph}}$, $u_{h_{\pm}}=\frac{1}{r_{h_{\pm}}}$ and $u_{pe}=\frac{1}{r_{pe}}$ is the minimum positive root of $G\left(u\right)$.}\label{GU}
	\end{figure}
		
		In Fig.~\ref{GU}, we have plotted the function $G(u)$. From the figure, we observe that for $b\geqslant b_c$, the behavior of photons in the quantum modified spacetime is similar to that in the Schwarzschild spacetime. However, for $b<b_c$, photons penetrate the outer horizon of the QBH, entering the inner horizon and reaching its perihelion within it. In fact, according to LQG theory, after reaching $r_{pe}$, photons will enter another universe and emerge from a WH in that universe. The trajectory of the photon after emerging from the WH will be completely symmetric to its previous trajectory. The additional accretion disk imaging caused by the WH results from the contribution of these photons.
		\newpage
		According to the above discussion, for photons with $b>b_c$ and $b<b_c$, their total deflection angle $\varphi$ in spacetime is
		\begin{equation}\label{eq2_17}
			\varphi=2\int_{0}^{u_{pe}}\frac{1}{\sqrt{G\left(u\right)}}du.
		\end{equation}
		For photon corresponding to $b=b_c$, it will arrive at $r_{ph}$ and then perpetually undergo circular motion.
		
\begin{figure}[htbp]
			\centering
			\includegraphics[width=1\textwidth]{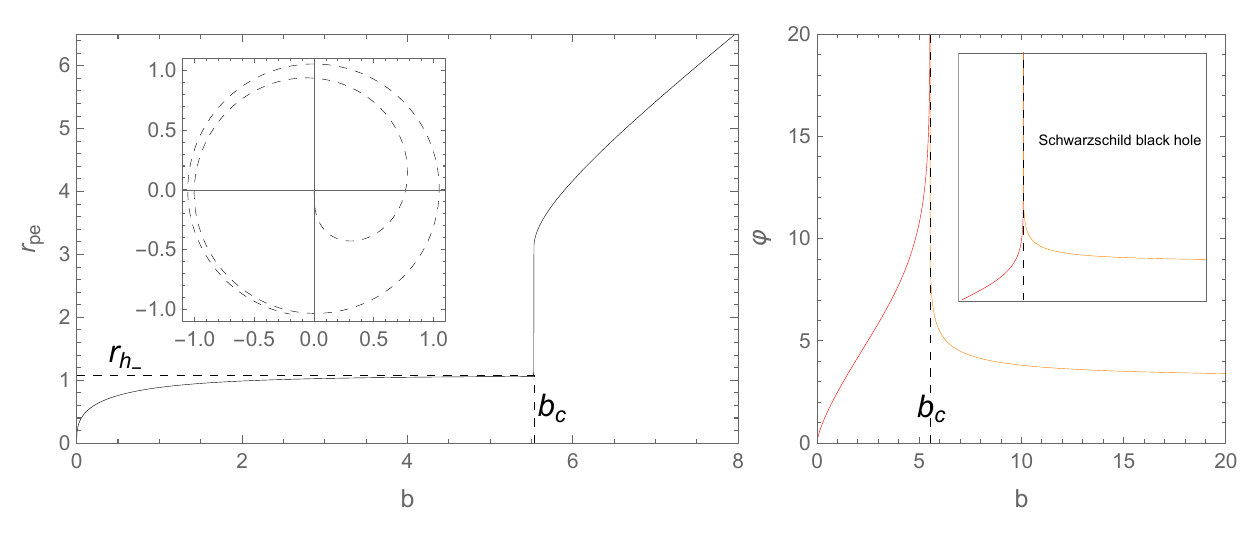}
			\caption{The left side and right side are variations of $r_{pe}$ and $\varphi$ with respect to $b$ respectively.}\label{jrd}
			\includegraphics[width=1\textwidth]{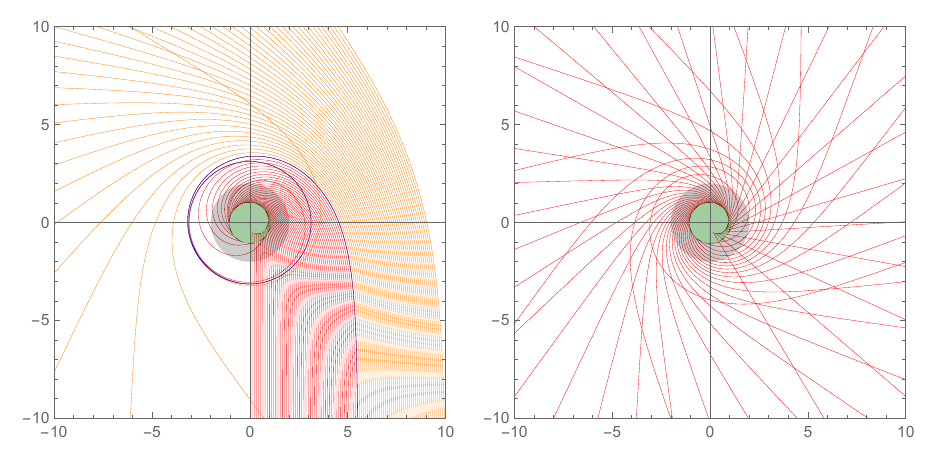}
			\caption{The left side and right side are the trajectories of photons in the spacetime of the BH$^\prime$ and the spacetime of the WH  respectively. The red, blue and orange curves represent the photons which meet $b<b_{c}$, $b=b_{c}$ and $b>b_{c}$ respectively.}\label{GL}
		\end{figure}

		In Fig.~\ref{jrd} on the left side, we illustrate $r_{pe}$ with respect to $b$, which consistently remain smaller than the radius of the inner horizon $r_{h_{-}}$, and exhibit a mutation at point $b_c$. In the inset, we depict the trajectories of these perihelion curve in spacetime. On the right side of Fig
		.~\ref{jrd}, we plot the total deflection angle $\varphi$ of photons moving in quantum modified spacetime. For photons with $b<b_c$, due to their trajectory allowing them to emerge from the WH after entering the BH$^\prime$, their total deflection angle is greater compared to that in Schwarzschild spacetime.
		
		Using numerical method, we have plotted the trajectories of photon motion in Fig.~\ref{GL}. The larger gray disk and the smaller green disk represent the region enclosed by the outer horizon $r_{h_{+}}$ and the inner horizon $r_{h_{-}}$ respectively. The spiral line within the green region is perihelion curve with $b<b_c$. In the left figure, photons with $b<b_c$ enter the BH$^\prime$ and emerge from the WH in the right figure.
	\newpage
	\section{Image of thin accretion disk for BH and WH}\label{sec3}
	\subsection{Observation coordinate system}\label{sec3_1}
	
	\begin{figure}
		\centering
		\subfigure{
			\includegraphics[width=1\textwidth]{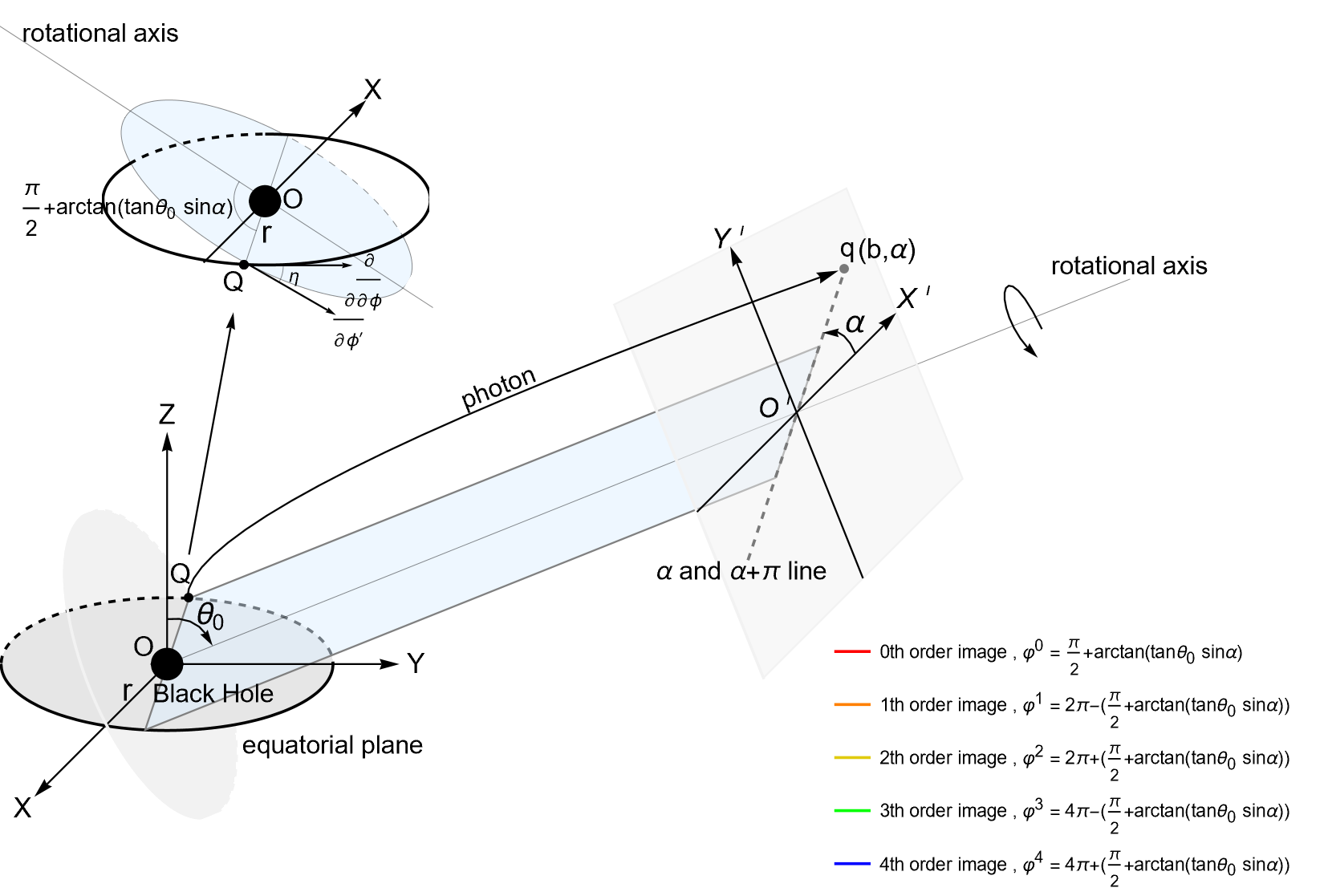}}
		\subfigure{
		\includegraphics[width=1\textwidth]{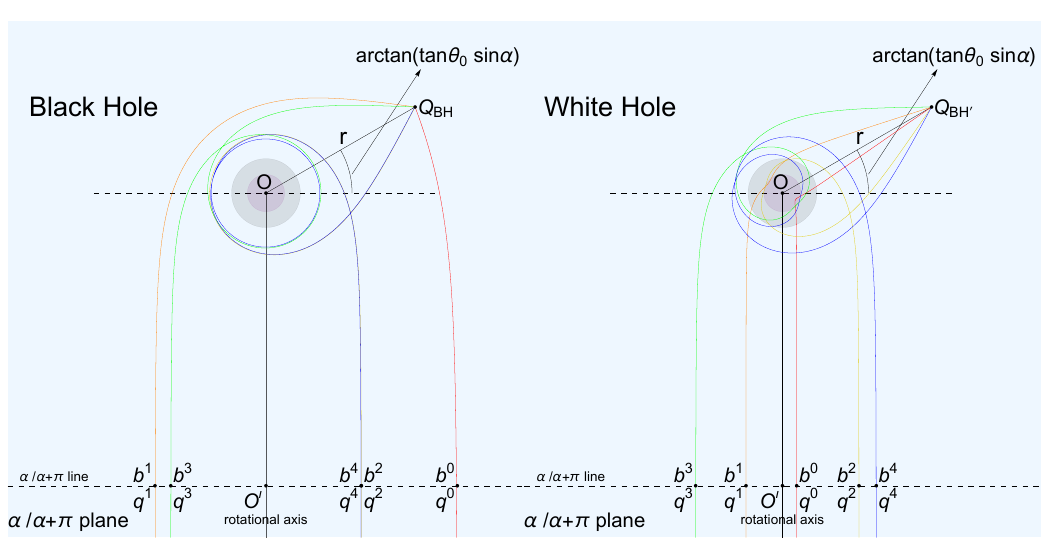}}
		\caption{Coordinate system.}\label{fig_coordinate}
	\end{figure}
	    Due to the independence of two universes, we can always build their coordinate systems to ensure a photon's null-geodesics in two universes smoothly link together at their perihelions. 
	    
	    To investigate the image of thin accretion disk, the observation coordinate system is shown as Fig.~\ref{fig_coordinate}. The observer locates at $\left(\infty,\theta_0,0\right)$ in the QBH's spherical coordinate system $\left(r,\theta,\phi\right)$, which takes center of QBH as $r=0$. Consider in the observer coordinate system $O'X'Y'$, a photon departing from the point $q\left(b,\alpha\right)$ in the vertical direction, which means that $b$ is precisely photon's impact parameter. This photon arrives at a point $Q\left(r,\frac{\pi}{2},\phi\right)$ in accretion disk of BH or BH$^\prime$. Because of the reversibility of light path, the photon from $Q\left(r,\frac{\pi}{2},\phi\right)$ will finally reach the its image point $q\left(b,\alpha\right)$. 
		\par
		If we fix $r$, we will derive the image of equal-$r$ orbit. As shown in the left side of Fig.~\ref{fig_coordinate}, every $\alpha/\alpha+\pi$ plane and equal-$r$ orbit in equatorial plane have two intersection points with the difference of azimuth angel $\phi$ being $\pi$. We set $\alpha=0$ and $\phi=0$ for $X'$-axis and $X$-axis respectively. According to geometry, the angle $\varphi$ between rotational axis and $OQ$ is~\cite{YOU}
	\begin{equation}\label{eq3_1}
		\varphi=\frac{\pi}{2}+\arctan\left(\tan\theta\sin\alpha\right).
	\end{equation}
		When $b$ gets closer to $b_{c}$, the light will bend more severely. So a source point $Q$ could have many image points $q$. We label these image points according to their $\varphi$ from small to large as $q^{n}$ ($n\in \mathbb{N}$), which represents $n^{th}$-order image respectively. As shown in the right side of Fig.~\ref{fig_coordinate}, all the even-order images of $Q$ are in the same side ($\alpha$) as $Q$. In contrast, all the odd-order images of $Q$ will appear in the opposite side ($\alpha+\pi$). Furthermore, the higher BH image's order is, the smaller its corresponding $b$ will become. And condition is opposite in WH. We will delve into this point in detail later on. We mark photon's changes of $\phi$ resulting in the $n^{\rm{th}}$-order image as $\varphi^{n}$: 
	\begin{equation}\label{eq3_2}
		\varphi ^n=
		\begin{cases}
		\frac{n}{2} 2\pi +\left( -1 \right) ^n\left[ \frac{\pi}{2}+\mathrm{arc}\tan \left( \tan \theta \sin \alpha \right) \right]&,\mathrm{when}~n~\mathrm{is~ even},\\
		\frac{n+1}{2} 2\pi +\left( -1 \right) ^n\left[ \frac{\pi}{2}+\mathrm{arc}\tan \left( \tan \theta \sin \alpha \right) \right]&,{\mathrm{when}~n~\mathrm{is~ odd}}.
		\end{cases}		
	\end{equation}
	 	Substituting these $\varphi^{n}$ into Eq.~\ref{eq2_16} one can get the their corresponding impact parameters $b^{n}$. The image point of source point $Q$ in observer coordinate system $O'X'Y'$ can be expressed as $q^{n}\left(b^{n},\alpha\right)$ for even number $n$ and $q^{n}\left(b^{n},\alpha+\pi\right)$ for odd number $n$.
	\subsection{Image of equal-$r$ orbit on thin accretion disk}\label{sec3_2}
\begin{figure}[h]
			\centering
			\includegraphics[width=1\textwidth]{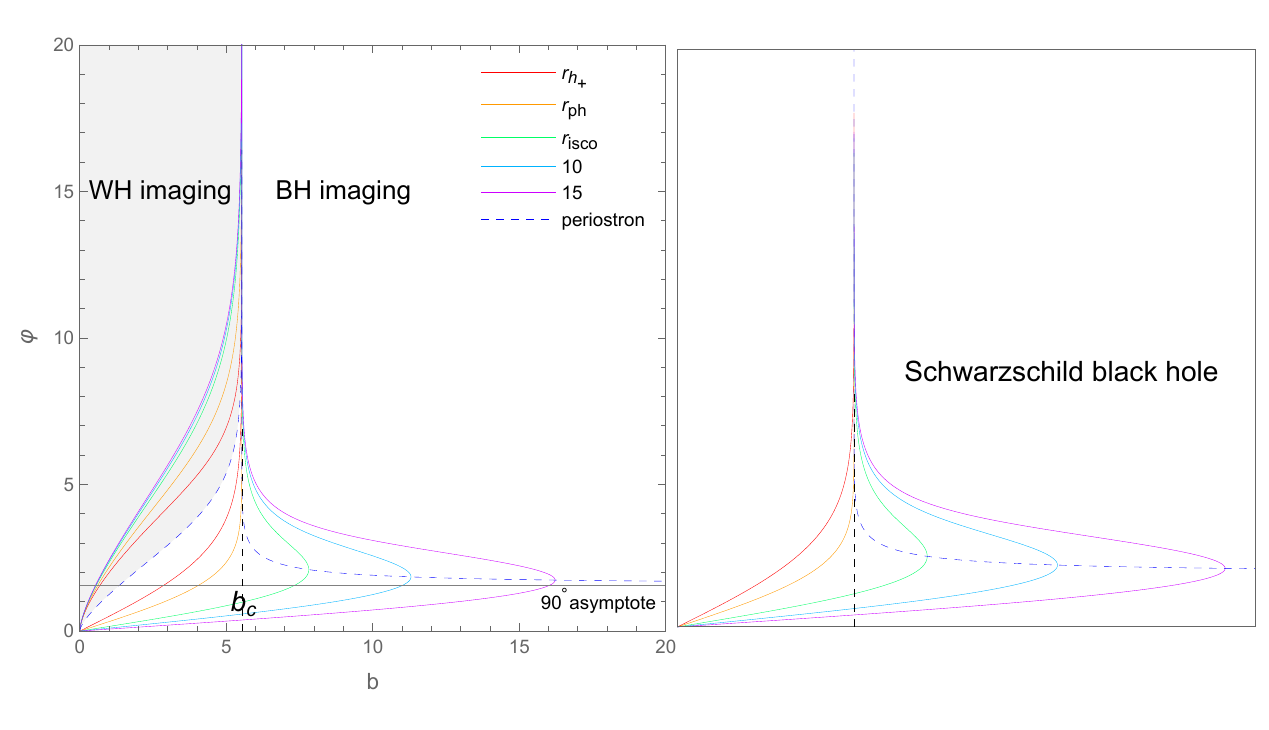}
			\caption{Deflection angle $\varphi\left(b\right)$ corresponding to intersections as a function of $b$ for different $r$. The left and right side are the QBH and SBH spacetimes respectively. $r_{isco}$ is radius of the innermost stable cicular orbit(ISCO) of time-like particle.}\label{fig_rphib}
		\end{figure}

		For photons coming form infinity with different values of $b$ on their trajectory plane, they will have different intersections with equal-$r$ orbit.
		Fig.~\ref{fig_rphib} gives figure of $\varphi\left(b\right)$. We denote the blue dashed line as $\varphi_{blue}(b)$. Using the blue dashed line as the demarcation, we designate the colored curves below the demarcation line as $\varphi _{color-down}(b)$ and those above it as $\varphi _{color-up}(b)$.
		\begin{equation}
			\varphi_{blue} \left( b \right) =\int_0^{u_{pe}}{\frac{1}{\sqrt{\frac{1}{b^2}-f\left( \frac{1}{u} \right)}}}du
		\end{equation}
		
		\begin{equation}\label{cd}
			\varphi _{color-down}\left( b \right) =\int_0^{u_{r}}{\frac{1}{\sqrt{\frac{1}{b^2}-f\left( \frac{1}{u} \right)}}}du
		\end{equation}
		
		\begin{equation}
			\begin{split}
				\varphi _{color-up}\left( b \right) &=2\int_0^{u_{pe}}{\frac{1}{\sqrt{\frac{1}{b^2}-f\left( \frac{1}{u} \right)}}}du-\int_0^{u_r}{\frac{1}{\sqrt{\frac{1}{b^2}-f\left( \frac{1}{u} \right)}}}du\\
				&=2\varphi_{blue}(b)-\varphi _{color-down}(b)
			\end{split}
		\end{equation}
		
		In the figures, every colored line represents a equal-$r$ orbit, point $ \left(b, \varphi \right)$ in colored line indicates that the deflection angel when photon that takes $b$ as its impact parameter arrives at equal-$r$ is $\varphi$. Blue dashed line will intersect with colored lines at their peaks, meaning that the point $\left(b, \varphi \right)$ on blue dashed line represents that the deflection angel when photon that takes $b$ as its impact parameter arrives at its perihelion $r_{pe}$. The trajectories within the shaded region represent equal-$r$ orbit near another BH in a separate universe. The colored lines within the gray region correspond to the orbits of the accretion disk around the BH$^\prime$, with the WH transporting photons with $b<b_c$ from these orbits to our universe. The remaining colored lines correspond to the orbits of the accretion disk around the BH in our universe.
	
\begin{figure}[htbp]
			\centering
			\includegraphics[width=1\textwidth]{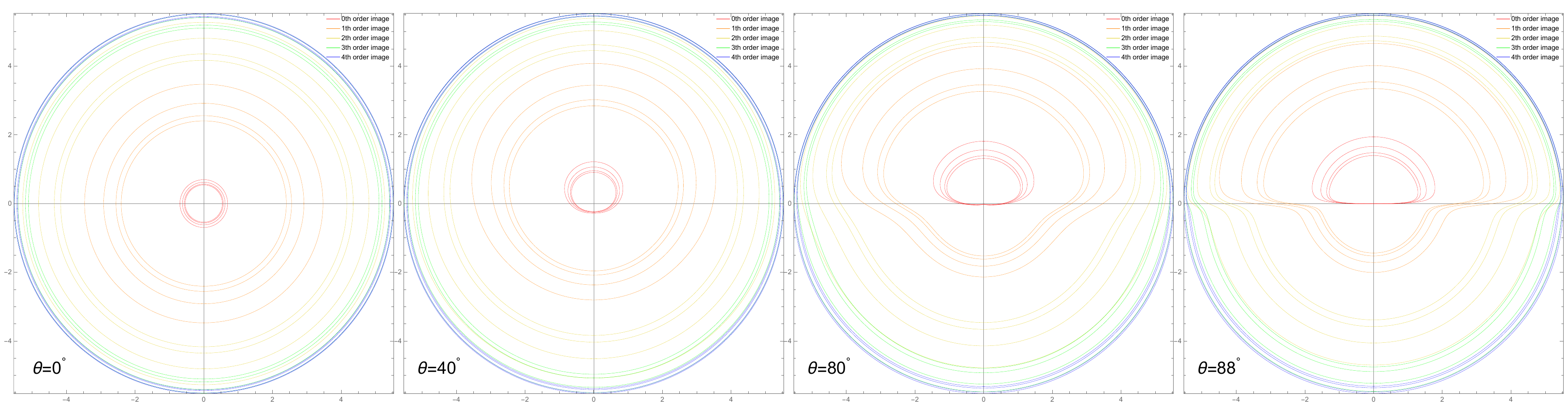}
			\caption{The images of the equal-$r$ orbit around the BH$^\prime$ transported by the WH.}
			\label{wh}
		\end{figure}

		To investigate the image of equal-$r$ orbit, we plot inverse function $b\left(\varphi\right)$ and $\alpha\left(\varphi\right)$ as Fig.~\ref{fig_1bphi}. One could clearly see in the figures that behaviours of $b\left(\varphi\right)$ for WH is similar with the condition when $r<r_{ph}$ and contrary to some cases of $r>r_{ph}$ of BH imaging. We take $\theta=\ang{0}$, $\ang{40}$, $\ang{80}$, $\ang{88}$ for an example and give these following analyses (we focus our analysis on the case of the WH):
		\begin{figure}[htbp]
			\centering
			\includegraphics[width=1\textwidth]{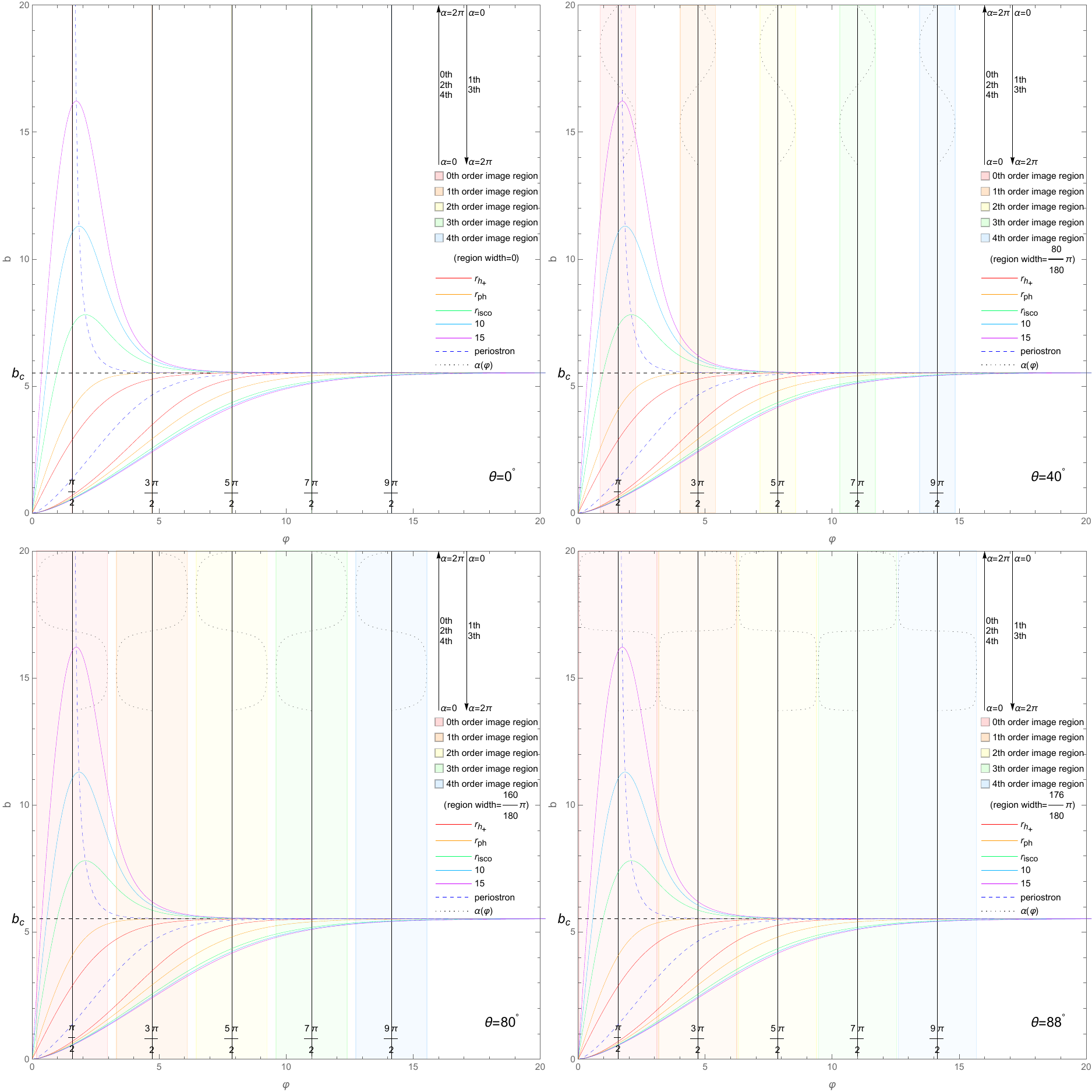}
			\caption{Figure of $b\left(\varphi\right)$. S-shaped curves are  inverse functions of Eq.~\ref{eq3_2} corresponding to respective images. For every even-order image, $\alpha$ takes $0$ to $2\pi$ form bottom to top. And odd-order image is opposite.}\label{fig_1bphi}
		\end{figure}
		\newpage
		\begin{itemize}
		\item$\theta=\ang{0}$
		\begin{itemize}
		\item[(a)] For any $\alpha$, $\varphi$ will be always $\frac{\pi}{2}$, which means the image of equal-$r$ orbit is a perfect circle. Contrary to the case of BH, as $r$ increases, the radius of the image $b$ decreases. Reflected to the Fig.~\ref{wh}, for the orbits with larger radii, their images will instead be located more inward;
		\item[(b)] With the increase of images' order $n$, $\Delta b$ caused by two different values of $r$ will increase initially and then decrease, and arrive at its maximum $\Delta b_{\rm{max}}$ near $n=2$. Furthermore, when $n\geq 4$, the images become hard to distinguish. Moreover, in contrast to the condition of BH, for a fixed $n$, as $r$ increases, different colored line will become increasingly crowded together, implying that orbits with larger radii will become difficult to distinguish;
		\item[(c)] The maximum $b_{\rm{max}}$ of colored lines (equal-$r$ orbit) will increase as $n$ increases, and $b_{\rm{max}}$ will take $b_{c}$ as its limit as $n$ increases.
		\end{itemize}
		\item $\theta=\ang{40}$, $\ang{80}$, $\ang{88}$
		\begin{itemize}
		\item [(a)] Due to nonzero inclination angle, the interval of photon's deflection angle $\varphi$ has a width, which leads to that its image is no longer a circle. But it can be seen in the figure that image is symmetric about Y$'$-axis resulted by S-shaped curves;
		\item [(b)] Similar to the case for $\theta=0$, the $n^{\rm{th}}$ images are hardly distinguishable for $n\geq 4$. But compared to $\theta=0$, for images of each order, $\Delta b$ is no longer a constant;
		\item [(c)] For a fixed $n^{\rm{th}}$ image, $b$ will always achieve its maximum and minimum at $\alpha=\frac{\pi}{2}$ and $\alpha=\frac{3\pi}{2}$. The behaviour of $b_{\rm{max}}$ of colored lines (equal-$r$ orbit) is similar with the condition $\theta=0$.
	
		\end{itemize}
		\end{itemize}

\begin{figure}[htbp]
			\centering
			\includegraphics[width=1\textwidth]{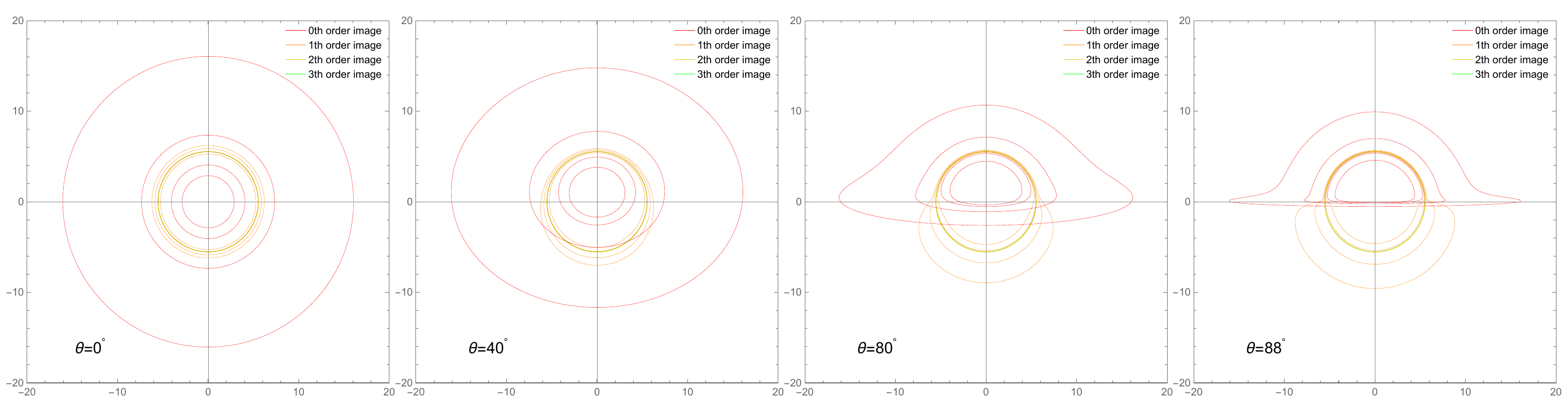}
			\caption{The equal-$r$ orbit images of BH.}
			\label{bh}
		\end{figure}

\begin{figure}[htbp]
		\centering
		\includegraphics[width=1\textwidth]{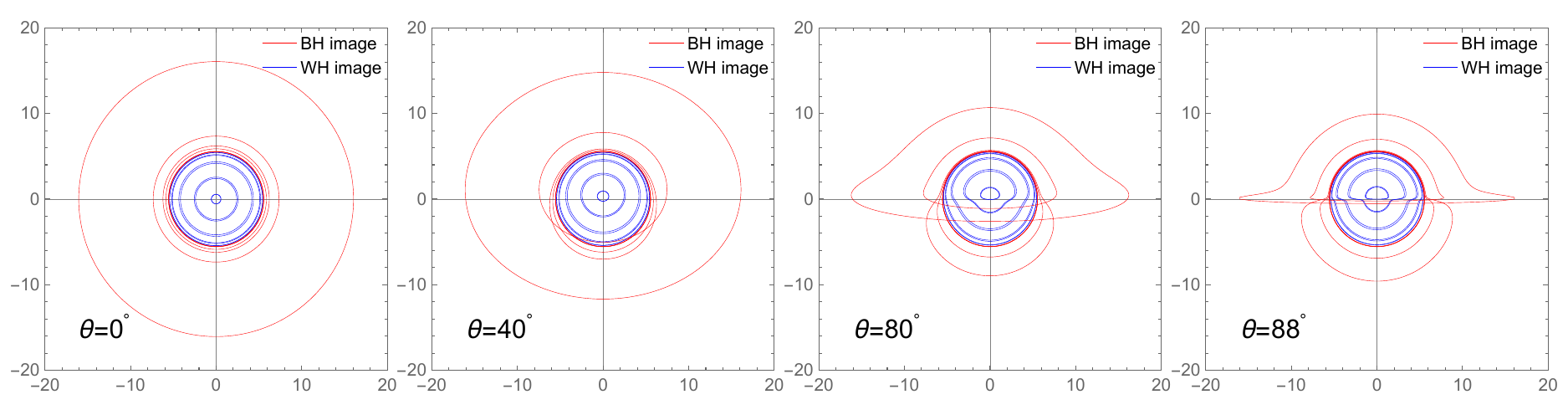}
		\caption{The merged equal-$r$ orbit images.}
		\label{bhwh}
	\end{figure}

		Now, in Fig.~\ref{bh}, we have plotted the images of equal-$r$ orbit around the BH; In Fig.~\ref{wh}, we have plotted the images of equal-$r$ orbit caused by the WH; Although BH and WH appear sequentially during the collapse of a dust ball, we have still merged their corresponding images in Fig.~\ref{bhwh} for ease of comparison.
		
	\subsection{Light intensity distribution on accretion disk}\label{sec3_3}
		We use Novikov-Thorne model to study the image of thin accretion disk with brightness \cite{NT}:
	\begin{equation}\label{eq3_3}
		F_{em}\left(r\right)=-\frac{\mathcal{M}\Omega'}{4\pi\sqrt{-g}\left(E-\Omega L\right)^{2}}\int_{r_{isco}}^{r} \left(E-\Omega L\right)L'dr,
	\end{equation}
		where $F(r)$ is radiation flux emitted by time-like particle in circular orbit $r$, $\mathcal{M}$ is black hole's accretion rate, $g$ is metric
		determinant, and $r_{isco}$ is radius of ISCO we have introduced previously, $E$, $\Omega$, and $L$ are energy, angular velocity, and angular momentum of time-particle moving on the disk.
		\par
		Due to the different gravitational fields at disk and observer and their relative motion, the frequency shift will happen, the radiation flux observer received will be \cite{HY,F1,F2}
	\begin{equation}\label{eq3_4}
		F_{obs}=\frac{F_{em}\left(r\right)}{\left(1+z\right)^{4}},
	\end{equation}
		where $z$ is redshift factor. From~\cite{YOU}, we obtain
	\begin{equation}\label{eq3_17}
		F_{obs}=\frac{-\frac{\mathcal{M}\Omega'}{4\pi\sqrt{-g}\left(E-\Omega L\right)^{2}}\int_{r_{isco}}^{r} \left(E-\Omega L\right)L'dr}{\left(\frac{1+\Omega b \sin\theta\cos\alpha}{\sqrt{-g_{tt}-g_{\phi\phi}\Omega^{2}}}\right)^4},
	\end{equation}	
	where
	\begin{equation}\label{eq3_19}
		\Omega=\pm\sqrt{-\frac{g_{tt}'}{g_{\phi\phi}'}},
	\end{equation} 
	\begin{equation}\label{eq3_20}
		E=-g_{t\mu}U^{\mu}=\frac{-g_{tt}}{\sqrt{-g_{tt}-g_{\phi\phi}\Omega^{2}}},
	\end{equation}
	\begin{equation}\label{eq3_21}
		L=g_{\phi\mu}U^{\mu}=\frac{g_{\phi\phi}\Omega}{\sqrt{-g_{tt}-g_{\phi\phi}\Omega^{2}}}.
	\end{equation}
	
\begin{figure}[h]
		\centering
		\includegraphics[width=1\textwidth]{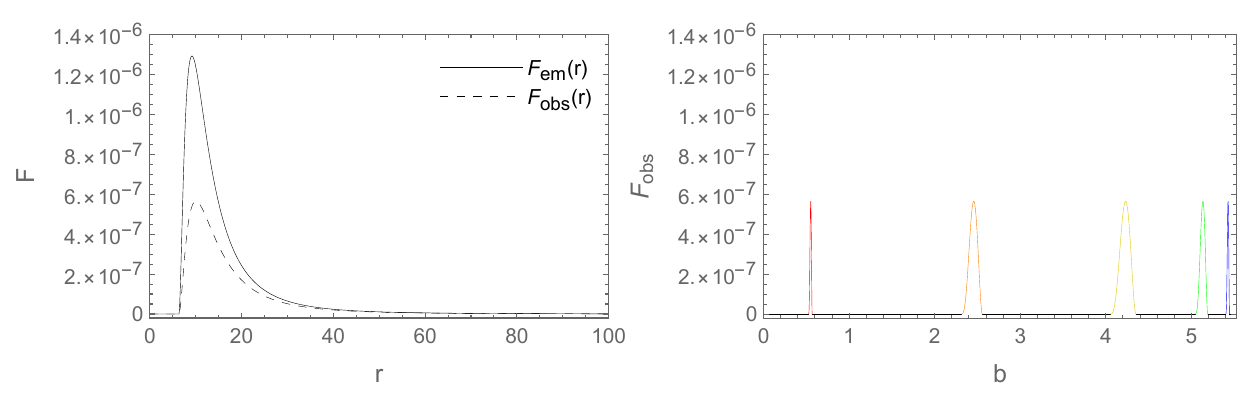}
		\caption{The left side is comparison of radiation flux at light source and observer. The right figure gives the first five order of observable radiation flux of WH's image as functions of $b$. We set $\theta=0$.}\label{fig_Feo}
	\end{figure}

	We give the figures of radiation flux $F\left(r\right)$ as Fig.~\ref{fig_Feo}. From the figure, we can observe that the gravitational redshift effect causes a decrease in the intensity of light received by the observer. In the figure on the right, from left to right, we observe the distribution of intensity for WH images' order 0 to 4. Among these, the distribution of the second order is the widest, as also evident from our previous Fig.~\ref{fig_1bphi}. Beyond the fourth order, the resolution becomes almost indistinguishable.
	
	In Tab.~\ref{bg}, we present the numerical values of the positions and widths of the first five-order images for different $M$ when $\theta_{0}=\ang{0}$. $b_{min}$ (corresponding $r=\infty$) and $b_{max}$ (corresponding $r=r_{isco}$) respectively represent the inner radius and outer radius of each order's corresponding annulus, while $\Delta b$ denotes the width of the annulus.
	
\begin{figure}[htbp]
		\centering
		\subfigure{
			\includegraphics[width=1\textwidth]{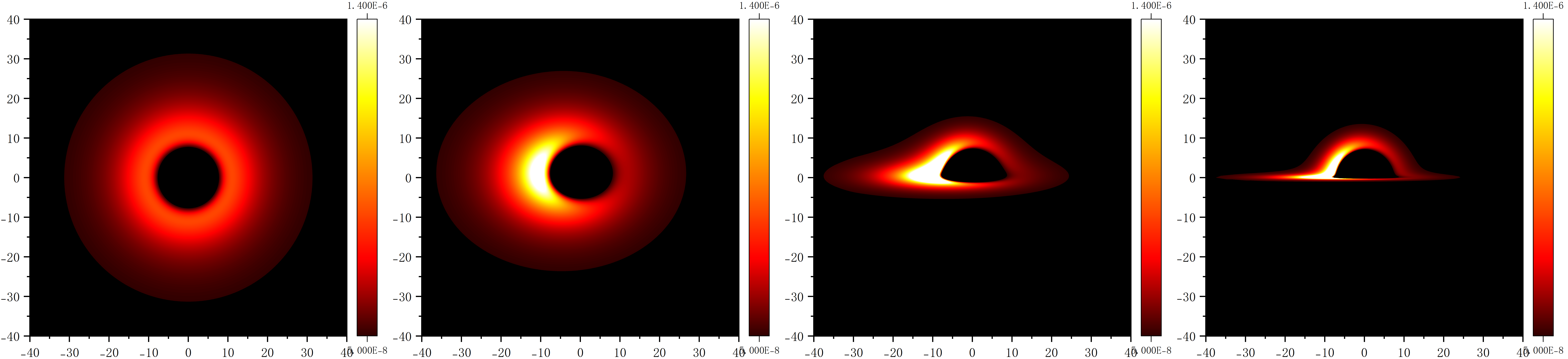}}
		\subfigure{
			\includegraphics[width=1\textwidth]{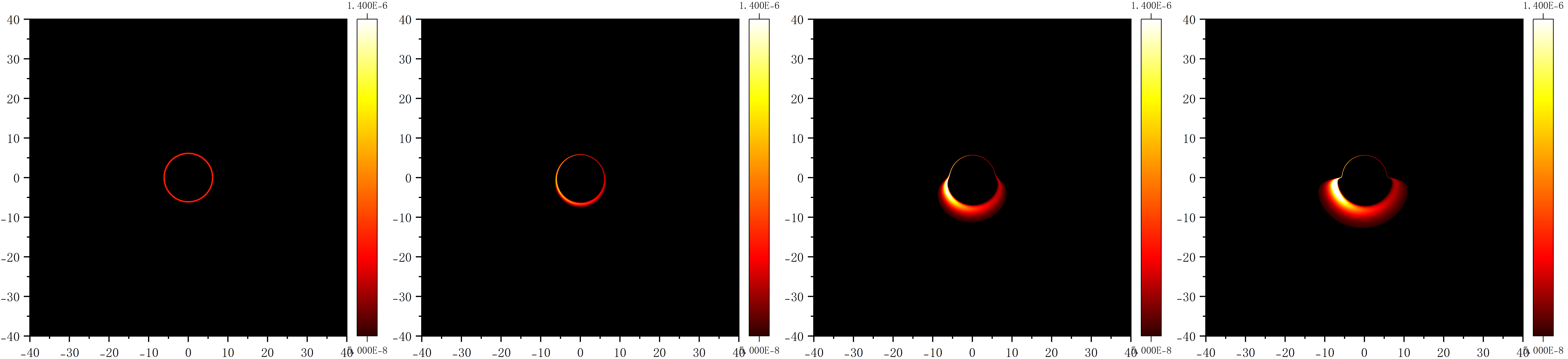}}
		\caption{The intensity images of the accretion disk around the BH. From left to right, they correspond to $\theta=\ang{0}$, $\ang{40}$, $\ang{80}$, and $\ang{88}$. The first row represents $0^{\rm{th}}$ images, while the second row represents $1^{\rm{st}}$ images.}
		\label{fig_4intensity}
		\includegraphics[width=1\textwidth]{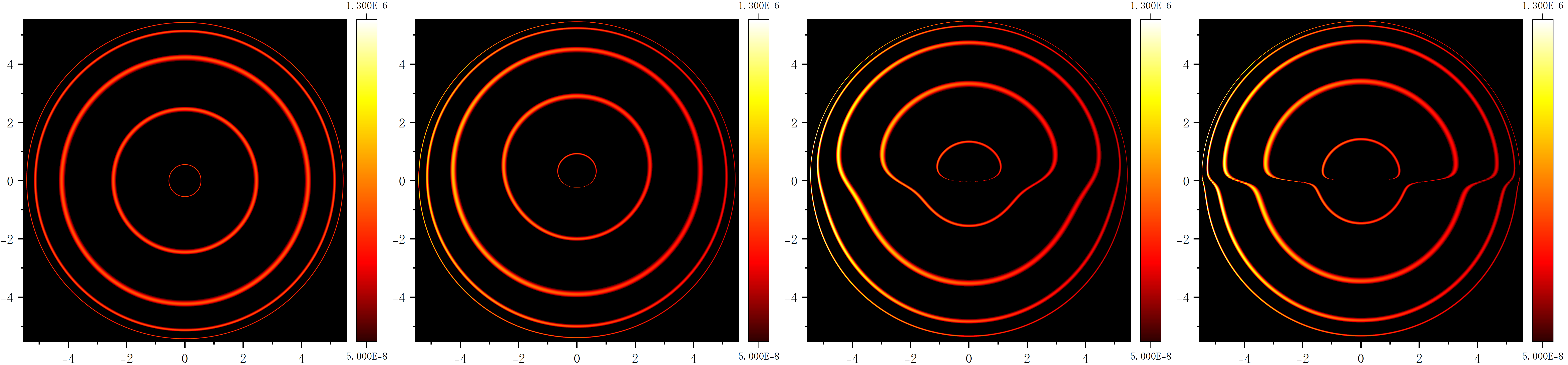}
		\caption{The intensity imaging of the accretion disk caused by the WH. From left to right, they correspond to to $\theta=\ang{0}$, $\ang{40}$, $\ang{80}$, and $\ang{88}$. From innermost to outermost, they represent $0^{\rm{th}}$, $1^{\rm{st}}$, $2^{\rm{nd}}$, $3^{\rm{rd}}$, and $4^{\rm{th}}$ images.}\label{WHGQ}
	\end{figure}

	According to the analysis above, in Fig.~\ref{fig_4intensity}, we depict the intensity distribution of the accretion disk around a BH, while in Fig.~\ref{WHGQ}, we illustrate the intensity distribution caused by WH. It can be observed from the figures that with increasing viewing angle, the Doppler effect becomes more pronounced, resulting in an asymmetrical distribution of the accretion disk's intensity. Furthermore, in our model, the effect of WH on the intensity is identical to that of BH.
	
	\begin{table}[htbp]
		\centering
		\includegraphics[width=0.45\textwidth]{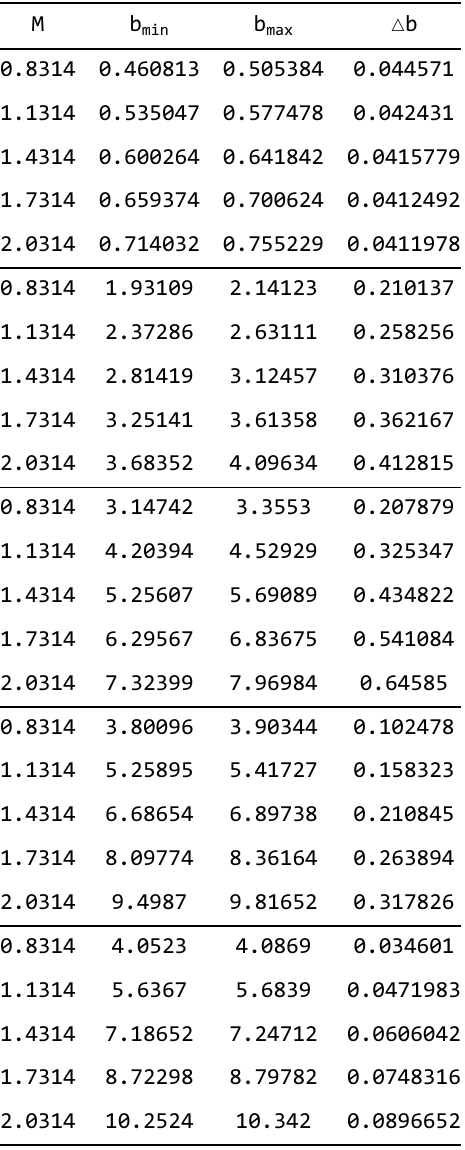}
		\caption{When $\theta_{0}=\ang{0}$,the positions and widths of the first five-order images for different $M$. From top to bottom, they represent $0^{\rm{th}}$, $1^{\rm{st}}$, $2^{\rm{nd}}$, $3^{\rm{rd}}$, and $4^{\rm{th}}$ images.}\label{bg}
	\end{table}
	
	\newpage
	\section{SUMMARY AND DISCUSSION}\label{sec4}
	Our work has led to the imaging of accretion disk caused by the WH. The results showed that the mechanism of WH and BH imaging are similar inside the photon sphere and opposite in some of cases outside. The imaging of accretion disk caused by the WH is entirely confined within a circle of radius $b_c$, as only photons with $b<b_c$ can be absorbed by the BH$^\prime$ and then emerge from the WH. We know that for BH, only the first two orders of images have high resolution, but for WH, the images caused by them becomes difficult to resolve only after the fourth order of images. In our QBH model, the resolution of the second order of images is the highest. In the case of BH, the images of accretion disk orbits with larger radii are located further outward, but in the case of WH, the images of accretion disk orbits with larger radii are located inward. Furthermore, as $r$ increases, the resolution of the images also decreases. With increasing viewing angles, the lower part of the images starts to diminish. In terms of intensity, the behavior of BH and WH on the intensity of light is the same. 
	
	In the model proposed by Lewandowski et al., a collapsing dust ball gradually evolves from a BH to a WH. If LQG theory is correct, then we will first observe the images of accretion disk in the BH phase, followed by the images of accretion disk in the WH phase. The images of accretion disk in these two phases is completely different, making it very easy for observers to distinguish such changes during observations. This will provide valuable references for astronomical observations to validate QG theory.
	
	\section*{Conflicts of interest}
	The authors declare that there are no conflicts of interest regarding the publication of this paper.
	\section*{Acknowledgments}
	We are grateful to Jie-Shi Ma and Yu-Cheng Tang for their useful suggestions. We also thank the National Natural Science Foundation of China (Grant No.11571342) for supporting us on this work.

	\bibliographystyle{unsrt}
	\bibliography{refff.bib}
\end{document}